# Proposal for a United Nations Basic Space Technology Initiative

Werner R. Balogh and Hans J. Haubold

*Office for Outer Space Affairs, United Nations, Vienna International Centre, P.O. Box 500, 1400 Vienna, Austria*

werner.balogh@unoosa.org, hans.haubold@unoosa.org

**Abstract**

The United Nations Programme on Space Applications, implemented by the United Nations Office for Outer Space Affairs, promotes the benefits of space-based solutions for sustainable economic and social development. The Programme assists Member States of the United Nations to establish indigenous capacities for the use of space technology and its applications. In the past the Programme has primarily been focusing on the use of space applications and on basic space science activities. However, in recent years there has been a strong interest in a growing number of space-using countries to build space technology capacities, for example, the ability to develop and operate small satellites. In reaction to this development, the United Nations in cooperation with the International Academy of Astronautics has been organizing annual workshops on small satellites in the service of developing countries. Space technology related issues have also been addressed as part of various other activities of the Programme on Space Applications. Building on these experiences, the Office for Outer Space Affairs is now considering the launch of a new initiative, preliminarily titled the United Nations Basic Space Technology Initiative (UNBSTI), to promote basic space technology development. The initiative would be implemented in the framework of the Programme on Space Applications and its aim would be to help building sustainable capacities for basic space technology education and development, thereby advancing the operational use of space technology and its applications.

## 1. Introduction

Following the launch of the first Earth-orbiting satellites in the late 1950's, the potential for space technology and its applications to make essential contributions to economic and social development was quickly recognized. In 1959 the United Nations Committee on the Peaceful Uses of Outer Space was established as a permanent body of the General Assembly of the United Nations to address issues arising to its Member States from space-related activities (United Nations, 1959). Based on recommendations of the first United Nations Conference on the Exploration and Peaceful Uses of Outer Space (UNISPACE), held in 1968, the United Nations Programme on Space Applications was launched in 1971. The Programme, implemented by the United Nations Office for Outer Space Affairs (UNOOSA), promotes the benefits of space-based solutions. It disseminates information on space activities, organizes a series of annual workshops, symposiums and expert meetings, and provides training opportunities in the practical applications of space technology (http://www.unoosa.org/oosa/en/sapidx.html).

Following the second United Nations Conference on the Exploration and Peaceful Uses of Outer Space (UNISPACE'82) held in 1982, the mandate of the Programme on Space Applications was expanded to include the promotion of the development of indigenous capabilities in the developing countries. It was acknowledged that opportunities for long-term intensive education at a higher level are a prerequisite for developing the skills and knowledge of university educators, engineers and research and application scientists. To address the lack of such education opportunities in space-related fields in the developing countries, four Regional Centres for Space Science and Technology Education, affiliated to the United Nations, were established in Africa (Centres in Morocco and Nigeria), Asia and the Pacific (Centre in India) and in Latin America and the Caribbean (Centre with campuses in Brazil and Mexico). The Centres provide specialized courses in rigorous theory, research and applications, as well as field exercises and pilot projects in those aspects of space science and technology that can contribute to sustainable development in each country (http://www.unoosa.org/oosa/en/SAP/centres/index.html).

Complementing its activities related to the use of space applications the Programme on Space Applications in 1991 launched a range of activities focussing on basic space science which were later integrated in the framework of the United Nations Basic Space Science Initiative (UNBSSI) (http://www.unoosa.org/oosa/en/SAP/bss/index.html) (Wamsteker et al, 2004). A total of twelve basic space science workshops were held from 1991 to 2004. From 2004 onwards the initiative has been continued through a series of annual workshops dedicated to the International Heliophysical Year of 2007 (IHY).

In summary it can be concluded that since its inception the focus of the activities of the United Nations Programme on Space Applications has been two-fold:

(i) Space applications, including remote sensing and geographical information systems, satellite communications and navigation, satellite meteorology and atmospheric sciences; and

(ii) Basic space science, including astronomy and astrophysics, solar-terrestrial interactions, planetary and atmospheric studies and exobiology.

Space applications and to a certain extent basic space science are based on the utilisation of space technology, for example to gather the necessary data and information from and in outer space. Given the wide and often free availability of space data today it is in principle not necessary to develop capabilities in space technology development to benefit from space applications. However, obviously the mastery and a good understanding of the technology underlying these applications are a precondition for achieving a higher level of independency and can enable those that have acquired the means to develop basic space technology to transition from being passive to becoming more active space users, for example, by allowing them to become providers of space-based data and information.

Consequently, the extent to which space technology should play a role within the activities of the United Nations Programme on Space Applications, has been considered on several occasions in the past

**2. Increasing interest in building capacity in space technology development**

In the 1990's and following discussions in preparation for and at the Third United Nations Conference on the Exploration and Peaceful Uses of Outer Space (UNISPACE III), held in Vienna, Austria, from 18 to 23 July 1999, capacity building in space technology development came increasingly into the focus of the United Nations Programme on Space Applications.

A workshop on the theme "Small satellites at the service of developing countries", organized by the Office for Outer Space Affairs in cooperation with the International Academy of Astronautics (IAA), was held in the framework of the Technical Forum of UNISPACE III (United Nations, 1999). Participants at the workshop concluded that small satellites were valuable tools in the development of a space infrastructure and scientific and application programmes, that they could play an important role in every country's space plan and offered opportunities for international cooperation. They also recommended that countries preparing a space plan should consider small satellites the most valuable tools to develop an indigenous space capability since they offered ideal opportunities for training and for learning the techniques associated with the design, development, manufacturing, testing and operation of a spacecraft.

In follow-up to this workshop, the Office for Outer Space Affairs and IAA and its Subcommittee on Small Satellites for Developing Nations continued to hold annual workshops on the same theme as part of the programme of the annual International Astronautical Congress. The recommendations and conclusions of these workshops have been published in a series of United Nations reports (see Table 1).

Workshop participants recognized that small satellites were a useful tool for the acquisition, development and use of space technology and for the associated development of a knowledge base and industrial capacity as well as for the development of spin-off enterprises. They noted the importance of space technology development in education curricula, especially for motivating and training students. In particular the development of small satellites could generate ingenious low-cost solutions, not only in individual educational endeavours, but also in more complex government-sponsored undertakings. Within a country, a small satellite programme could stimulate interest in science and technology, enhance quality of life and the quality of education, promote research and development and result in better linkages between government agencies, educational institutions and industries

Participants also concluded that small satellite technology could be used effectively in addressing regional problems and could promote international cooperation. Clear benefits could be derived from cooperation programmes, whether within a bilateral satellite programme or in a constellation of small satellites with individual satellites being provided by the cooperating countries. For example, a constellation of small Earth Observation satellites would not only result in improved temporal resolution but would also increase the operational stability of the observation system.

In summary, the participants concluded that there were tremendous spin-offs to be gained from introducing space activities through a small satellite programme. In view of these benefits they recommended that space activities should be an integral part of any national programme devoted to technology acquisition and development.

The growing interest in space technology development in general may in part also be attributable to the following trends:

o	The development of small satellites has become feasible for entities or organizations with comparatively small budgets and has relatively modest infrastructure requirements. For example, university institutes, including those in developing countries, have already launched or are in the process of developing small satellites or are considering doing so.

o	Information about space technology and its know how, specifically also related to small satellites, is increasingly becoming publicly available. The growth of the internet and the world wide web has contributed to facilitate the accessibility and sharing of this kind of information. The involvement of universities has also contributed to the increased sharing of information through academic publications and informal networks.

o	The development of standardized platforms and interfaces, such as the CanSat and CubeSat platforms, as well as the availability of standardized protocols, such as for telemetry, tacking and control of small satellites, has facilitated the entry of new players, including private entities, universities and institutions in small, formerly primarily space-using countries, to get involved with space technology.

o	A growing number of countries, including developing countries, have recently established or are in the process of establishing national space organizations and are contemplating national space programmes or seeking to strengthen their existing space activities,.

o	Other countries are procuring or considering to procure satellites on the commercial markets. An indigenous capacities in space technology development can help to assure that the best possible solution for the country is being procured.

o	The desire of space industry from established space nations to seek new markets in emerging space nations has also opened new venues for cooperation that may facilitate the development of indigenous competencies in space technology development or even the possibility of establishing a basic space industry capacity.

The establishment of a space industry capacity goes hand in hand with capacity building in space technology development.. The Programme on Space Applications has addressed issues related to the role of space industry in emerging space nations, including those in developing countries. A symposium on the theme "Space Industry Cooperation with the Developing World" was held in Graz, Austria, in 1997 (United Nations, 1997), (United Nations 1998). The growing role of the private sector in space activities was part of the discussions at UNISPACE III. As a result the Scientific and Technical Subcommittee of the Committee on the Peaceful Uses of Outer Space is organizing space industry symposiums at some of its annual sessions since 2000. The latest symposium in this series was held on 12 February

2008 and discussed the topic "Space industry in emerging space nations" (http://www.unoosa.org/oosa/COPUOS/stsc/2008/symposium.html). The discussions at the symposiums have confirmed the strong interest of countries to strengthen their space technology capabilities or even to develop basic space industry capacities.

The consistent recommendations of the annual United Nations/IAA workshops, the growing interest in establishing basic space technology development capacities and other trends outlined above, such as the fact that information and know-how about many of the relevant technologies have only recently been released into the public domain, suggest that this is the appropriate time to consider an initiative dedicated to space technology capacity building in the framework of the United Nations Programme on Space Applications. We therefore propose such an initiative, which we have preliminarily titled the United Nations Basic Space Technology Initiative (UNBSTI). The initiative would complement and complete the activities ongoing in the existing two pillars of the United Nations Programme on Space Applications in the fields of space applications and basic space science.

As suggested by its name, the proposed initiative would focus on basic space technology. While there is not yet a commonly agreed-upon definition of the term "basic space technology", it is understood by the author's that it constitutes such affordable space technology that can be developed with relatively small teams and modest infrastructure requirements. At the same time the initiative would limit itself to discuss space technology about which information is already widely available in public sources or that the technology owners agree to share with others so that technology transfer considerations are being taken into account.

**3. Objectives of a United Nations Basic Space Technology Initiative**

To define the specific objectives of the initiative it is first necessary to identify its stakeholders. It is understood that stakeholders are potential contributors, cooperation partners and beneficiaries, and include relevant governmental and non-governmental entities and organizations in United Nations Member States and their representatives. The Office for Outer Space Affairs will draw on its cooperation experiences and lessons learned in past activities of the Programme on Space Applications. Among the traditional cooperation partners are national and regional space agencies, regional space cooperation mechanisms, and academic institutions, including the Regional Centres for Space Science and Technology Education, affiliated to the United Nations. Cooperation will play an important role, as there are already other ongoing space technology-related initiatives and the United Nations should focus on where it can make the greatest impact to promote capacity building in space technology development.

To help with the identification of the possible objectives of the initiative, the Office for Outer Space Affairs has widely disseminated a questionnaire and engaged in informal discussions with stakeholders. The questionnaire is available upon request from the authors and seeks feedback on the following questions:

(i)   Is there a role for the United Nations, through its Programme on Space Applications, to seek involvement with activities related to basic space technology?

(ii) What could be the value-adding role of the United Nations, taking into consideration other activities and initiatives already ongoing in the field?

(iii) Which activities should be at the focus of a space technology initiative?

The questionnaire also solicits interest in contributing to the proposed United Nations initiative and seeks to identify potential partnerships and cooperation opportunities.

Based on the informal discussions and the feedback received in reply to the questionnaire there appears to be a consensus that, the Programme on Space Applications can have a role to play in addressing capacity building in space technology development-related issues. The replies indicate a strong interest in the development and operation of small satellites. There is the view that the United Nations could act as an honest and independent interlocutor, information broker and interface between stakeholders, specifically between countries and entities that have already demonstrated experience with space technology and countries and entities seeking to establish basic space technology capacities. This may help to enable some of these countries and entities to shift their focus from being a passive user of space technology and its applications to becoming a more knowledgeable and skilful space actor and possibly even a developer of basic space technology. Stakeholders also expressed their hope, that the initiative would help them to improve their know-how and skills so that they would become preferred partners for space cooperation. Many of the respondents also indicated their support, by offering to co-organize or host meetings in the framework of the proposed initiative.

In developing the initiative special attention has to be paid that its objectives are well integrated with other elements of the United Nations Programme on Space Applications and take account of the highly diverse levels of space technology capabilities of the stakeholders. It can be expected that some of the stakeholders will initially require assistance with building the necessary expertise, such as educating a critical mass of space technology experts, while other stakeholders may already have a long history of experience with the development of space technology. The initiative will have to take this into account when defining efficient and effective objectives so that it can serve the broadest number of stakeholders possible without running the danger of loosing focus.

Based on these initial discussions and drawing from the recommendations of the United Nations and IAA workshops on small satellites at the service of developing countries, the following provisional objectives have been identified for the UNBSTI:

(i) Share and exchange information on theoretical and practical aspects of basic space technology, including the design and operation of small satellites, their subsystems, payloads and support infrastructure:

- Systems engineering and mission design
- Satellite bus design
- Satellite subsystems: propulsion, structure, power, thermal, attitude determination & control, telemetry, tracking & control (including

              communication, command and data handling)
- Payload-specific technologies
- Experience with designing and launching of small satellites, including assembly, test and integration

(ii)      Support and provide assistance to basic space technology development-related capacity building efforts

(iii)     Promote international cooperation in basic space technology development

(iv)     Make use of synergies with other activities of the United Nations Programme on Space Applications to address issues where basic space technology can provide specific solutions to address pressing problems

      Table 2 summarizes some of the potential contributions that stakeholders could make as well as the potential benefits they could derive from the initiative. The agreement on a final initial set of objectives will be among the first tasks of the proposed initiative. Another task that would have to be tackled early-on could be a discussion on the type of activities that could be subsumed under the term "basic space technology". Guidance could be provided by similar discussions on the scope of the term "basic space science" in the framework of the UNBSSI (United Nations, 1991).

**4. Proposed work programme**

      To implement the proposed objectives, the following elements for a work programme could be considered:

(i)       Dissemination and exchange of information and educational resources on basic space technology development through workshops, conferences, expert meetings, training courses, websites and publications.

(ii)      Development of an education curriculum on basic space technology development and related issues, including education material on systems engineering, mission design, satellites, their buses, subsystems and payloads and operational aspects. The education curriculums on satellite communications, meteorology, remote sensing and space science developed for use by the Regional Centres for Space Science and Technology Education, affiliated to the United Nations and other educational institutions could provide a model (http://www.unoosa.org/oosa/en/SAP/centres/index.html).

(iii)     Propose and support projects to promote and coordinate the development of science and application payloads that could be deployed on standardized satellite platforms. This may include coordinated measurement campaigns and network science building on the experience gained with ground-based projects implemented in the framework of the UNBSSI and IHY (Haubold, 2003).

(iv)     Support existing or planned standardization and inter-operability efforts for ground stations, telemetry protocols, software and hardware standards, by promoting their use in world wide space technology projects.

(v) Cooperate with ongoing initiatives related to the development of small satellites.

In addition to the activities listed above, some discussions have also started to explore a possible role of the United Nations Basic Space Technology Initiative in activities related to global space exploration initiatives and to the utilization of the International Space Station, specifically considering opportunities in support of basic space technology-related activities, which could include the provision /mediation of experiment flight opportunities and the testing and space qualification of soft- and hardware developed in the framework of the initiative.

The proposed activities are presently merely part of a wish list and it is expected that the ongoing consultations and the initial workshops and meetings planned in the framework of the initiative will establish a refined list of potential activities, which will be tailored to the specific needs of the stakeholders.

A major concern for the initiative is to ensure that there is no duplication of ongoing efforts and that indeed value is being created for the stakeholders through the involvement of the United Nations. For example, the large number of already ongoing and planned CubeSat projects (http://en.wikipedia.org/wiki/CubeSat) suggests that the focus of the initiative should not be purely on the development of small satellite platforms. However, the United Nations may play a value adding role in promulgating useful, standardized application payloads together with the appropriate requirements, to be flown on standardized satellite platforms to perform coordinated measurement campaigns. Such an activity could provide a meaningful mission objective for ongoing or planned CubeSat development programmes. The data and information gathered with the help of such payloads could then be exchanged and exploited by the involved parties. Possible measurement campaigns could include the collection of data from the Earth's magnetic field or from other properties of the near-Earth space environment. The payload operations could build on the ongoing coordination efforts aiming to build a worldwide network of CubeSat ground stations for telemetry, tracking and control, such as the Global Educational Network for Satellite Operations (http://www.genso.org/) implemented under the auspices of the International Space Education Board (ISEB) with CNES, CSA, ESA, JAXA and NASA as member space agencies.

The experience gained with similar, but ground-based projects implemented in the framework of the United Nations Basic Space Science Initiative and the International Heliophysical Year (IHY) could be useful in this context. They include the BSS TRIPOD (Telescopes, Observing, and Teaching), the IHY TRIPOD (Instrument Arrays, Data, and Teaching) and the IHY Distributed Instrument Programme consisting of 17 networks (United Nations, 2007). A possible example for a project that could be considered in the framework of UNBSTI would be the development of a small standardized scientific payload for the observation of space weather. A network of independent satellites carrying that payload could provide potentially useful in-situ data from many locations in near-Earth space, which could complement the data of the ground-based measurements.

**5. Implementation plan**

Capacity building activities need to be implemented and assessed over time and the United Nations Basic Space Technology Initiative is therefore envisioned to be a multi-year effort. While it is considered that the initiative is well within the present objectives and priority thematic areas of the Programme on Space Applications and covered by its mandate, new activity proposals are presented for approval to member States of the Committee on the Peaceful Uses of Outer Space. It is therefore planned to present the initiative as part of the annual plan of activities for the Programme on Space Applications at the forthcoming, forty-sixth Session of the Scientific and Technical Subcommittee of COPUOS, to be held in Vienna, Austria, in February 2009.

The Office for Outer Space Affairs plans to dedicate a session at the fifth United Nations/ESA/NASA/JAXA Workshop on Basic Space Science and IHY 2007 to be held in Jeju, Republic of Korea, in September 2009, to kick-off the launch of the United Nations Basic Space Technology Initiative. The first dedicated UNBSTI workshop could be held as early as in the second half of 2009.

**6. Conclusions**

The aim of this paper is to provide preliminary information on the proposal for an initiative dedicated to capacity building in space technology development in the framework of the United Nations Programme on Space Applications. The initiative, preliminarily called the United Nations Basic Space Technology Initiative, would fill a gap by complementing the presently ongoing activities in the fields of basic space science and space applications. The initiative builds on past experience and lessons-learned and takes account of the most recent developments in the space technology sector, the growing interest in developing countries to establish or strengthen indigenous space technology capacities and first feedback received from potential stakeholders.

As the planning stage of the initiative continues to evolve, the Office for Outer Space Affairs wishes to engage in a dialogue with potential stakeholders, interested parties and experts in the field of space technology development and involved in capacity building activities in governmental and non-governmental. The authors welcome feedback that may contribute to further refine and strengthen the proposal.

*Disclaimer: The views expressed herein are those of the authors and do not necessarily reflect the views of the United Nations.*

**References**

Haubold, H., "Promoting research and education in basic space science: the approach of the UN/ESA workshops", Space Policy 19, p. 215-219, 2003.

United Nations, General Assembly Resolution 1472(XIV), "International Co-operation in the Peaceful Uses of Outer Space", 12 December 1959.

United Nations, Report on the first UN/ESA Workshop on Basic Space Science, Bangalore, India, 1991, A/AC.105/489, 1991.

United Nations, Report on the United Nations/European Space Agency Symposium on Space Industry Cooperation with the Developing World, Graz, Austria, 8-11 September 1997, A/AC.105/683, 12 December 1997.

United Nations, "United Nations / European Space Agency Symposium on Space Industry Cooperation with the Developing World", Proceedings of the United Nations/European Space Agency Symposium, 1997, United Nations, Vienna, V.98-55293, August 1998.

United Nations, Report of the Third United Nations Conference on the Exploration and Peaceful Uses of Outer Space, Vienna, 19-30 July 1999, United Nations publication, Sales No. E.00.I.3, annex III, 1999.

United Nations, Report on the third United Nations/European Space Agency/National Aeronautics and Space Administration workshop on the International Heliophysical Year 2007 and basic space science, Tokyo, Japan, 18-22 June 2007, A/AC.105/902, 12 December 2007.

Wamsteker, W., Albrecht, R., Haubold, H.J. (Eds.), Developing Basic Space Science World-Wide – A Decade of UN/ESA Workshops, Kluwer Academic Publishers, The Netherlands, 2004.

Table 1 **The United Nations and the International Academy of Astronautics have held a series of Workshops on Small Satellites in the Service of Developing Countries.**

| Date | Location | Special theme | Report[1] |
|---|---|---|---|
| 5 October 2000 | Rio de Janeiro, Brazil | The Latin American Experience | A/AC.105/745 |
| 2 October 2001 | Toulouse, France | The African Perspective | A/AC.105/772 |
| 12 October 2002 | Houston, United States of America | Beyond Technology Transfer | A/AC.105/799 |
| 30 September 2003 | Bremen, Germany | A Contribution to Sustainable Development | A/AC.105/813 |
| 5 October 2004 | Vancouver, Canada | Current and Planned Small Satellite Programmes | A/AC.105/835 |
| 19 October 2005 | Fukuoka, Japan | - | A/AC.105/855 |
| 3 October 2006 | Valencia, Spain | - | A/AC.105/884 |
| 25 September 2007 | Hyderabad, India | - | A/AC.105/897 |
| 30 September 2008 | Glasgow, Scotland | - | To be published |

---

[1] The reports are available in all official languages of the United Nations from the Official Document System of the United Nations at http://documents.un.org/

Table 2 **Stakeholders of the United Nations Basic Space Technology Initiative and their potential contributions and benefits.**

| Stakeholders | Potential contributions | Potential benefits |
|---|---|---|
| Member States | - Provide overall guidance to the Programme on Space Applications and the direction of UNBSTI<br>- Nominate national experts and focal points<br>- Offer to host UNBSTI-dedicated meetings<br>- Provide resources and in-kind contributions | - Assistance with establishing or strengthening capacity in space technology development, possibly as part of an overall research and development strategy<br>- Access to compiled space technology-related information, including best practices<br>- Participation of national experts in UNBSTI-dedicated activities and meetings<br>- Identify cooperation opportunities |
| Academic and educational institutions, including the Regional Centres for Space Science and Technology Education, affiliated to the United Nations, and research centres | - Share information about ongoing and planned projects and programmes, know-how and cooperation opportunities<br>- Nominate experts and focal points<br>- Contribute to the development of a space technology education curriculum | - Assistance with launching space technology-related activities<br>- Access to information, know-how and cooperation opportunities and best practices<br>- Opportunities for sharing of facilities<br>- Use of the space technology education curriculum in implementing new or improving existing space technology-dedicated academic programmes |
| Space agencies | - Provide in-kind support, such as experts, training opportunities, know-how and access to facilities<br>- Offer launch opportunities | - Benefit from an international framework for space technology cooperation<br>- Opportunities for exchange with space technology experts from emerging space nations<br>- Identify cooperation opportunities |
| Space industry | - Inform about relevant space industry activities and commercially available products<br>- Provide in-kind support, such as spare parts and surplus hardware<br>- Provide cooperation opportunities, such as the joint development of systems or the transfer of technology<br>- Provide employment and training opportunities | - Opportunities to meet and address decision makers and key personnel involved in space technology-related activities from all over the world<br>- Access to a skilled workforce<br>- Identify cooperation opportunities<br>- Establish and build new business opportunities |
| Regional space cooperation organizations | - Provide a regional forum for coordination of UNBSTI-related activities<br>- Provide guidance and make recommendations on region-specific issues<br>- Co-organize regional UNBSTI-dedicated meetings | - Benefit from an international framework for space technology cooperation<br>- Synergies with the United Nations Programme on Space Applications<br>- Identify intra- and inter-regional cooperation opportunities |
| International space-related non-governmental organisations | - Provide a forum for UNBSTI-related activities<br>- Co-organize UNBSTI-dedicated meetings<br>- Make specific contributions | - Opportunities to meet and address decision makers and key personnel involved in space technology-related activities from all over the world<br>- Identify new members or member organizations |